\documentclass{mn2e}
\usepackage{graphicx}

\newcommand{\dmthi}{{\partial m\over \partial\theta_i}}
\newcommand{\dmthj}{{\partial m\over \partial\theta_j}}

\newcommand{\beq}{\begin{equation}}
\newcommand{\eeq}{\end{equation}}
\newcommand{\beqa}{\begin{eqnarray}}
\newcommand{\eeqa}{\end{eqnarray}}
\newcommand{\omk}{\Omega_k}
\newcommand{\om}{\Omega_m}
\newcommand{\Om}{\Omega_m}
\newcommand{\calm}{{\cal M}}
\newcommand{\dl}{\delta}
\newcommand{\aj}{AJ}
\newcommand{\apj}{ApJ}
\newcommand{\apjl}{ApJL}
\newcommand{\prd}{PRD}
\newcommand{\prl}{PRL}

\newcommand{\procspie}{Proc~ of SPIE}
\newcommand{\aap}{A\&A}
\def\la{\mathrel{\mathpalette\fun <}}
\def\ga{\mathrel{\mathpalette\fun >}}
\def\fun#1#2{\lower3.6pt\vbox{\baselineskip0pt\lineskip.9pt
  \ialign{$\mathsurround=0pt#1\hfil##\hfil$\crcr#2\crcr\sim\crcr}}}

\vspace{0.3in} 
\begin{document} 
\title[Systematic Uncertainties on Cosmological Parameters] 
{Effects of Systematic Uncertainties on the
Supernova Determination of Cosmological Parameters} 
\author[Alex G.~Kim, Eric V.~Linder, Ramon Miquel, Nick Mostek]
{Alex G.~Kim$^1$, Eric V.~Linder$^1$, Ramon Miquel$^1$\thanks{E-mail address: 
{\tt rmiquel@lbl.gov}},
 Nick Mostek$^2$\\
$^1$Lawrence Berkeley National Laboratory, Physics Division,
1 Cyclotron Road, Berkeley, CA 94720, USA \\
$^2$Indiana University, Department of Astronomy,
Swain West 319, Bloomington, IN 47405, USA}
\maketitle
\baselineskip=14pt
\begin{abstract} 
Mapping the recent expansion history of the universe offers the 
best hope for uncovering the characteristics of the dark energy 
believed to be responsible for the 
acceleration of the expansion.
In determining cosmological and dark-energy 
parameters to the percent level, systematic uncertainties impose a floor 
on the accuracy more severe than the statistical measurement precision.  
We delineate the categorization, simulation, and 
understanding required to bound systematics for the specific case of 
the Type Ia supernova method.  Using simulated data of forthcoming 
ground-based surveys and the proposed space-based SNAP mission 
we present Monte Carlo results for
the residual uncertainties on the cosmological parameter 
determination.  The tight systematics control with optical and 
near-infrared observations and the extended redshift reach 
allow a space survey to bound the systematics below 
0.02 magnitudes at $z=1.7$.  For a typical SNAP-like supernova survey, this keeps total 
errors within 15\% of the statistical values and provides estimation 
of $\Omega_m$ to 0.03, $w_0$ to 0.07, and $w'$ to 0.3; these can 
be further improved by incorporating complementary data. 
\end{abstract} 
\section{Introduction} 
With the great increase in observational capabilities in the past and 
next few years, we can look forward to 
cosmological data of unprecedented volume and quality.  
These will be brought to bear on the outstanding 
questions of our ``preposterous universe'' \cite{ref:carroll} --what
new forms of matter and energy constitute 95\% of 
the universe? what is the underlying
nature of the mysterious dark energy causing the observed acceleration of 
the expansion of the universe yet without an explanation within the 
standard model of particle physics?  But 
more photons of any given observational method 
will not teach us the properties of the 
cosmological model before we understand the sources and intervening 
medium. Systematic uncertainties, rather than statistical errors, 
will bound our progress at the level where we fail to correct for 
astrophysical interference to our inference. 

This applies to each one of the promising cosmological probes. Use of the 
cosmic microwave background radiation has been dramatically successful 
in fitting certain cosmological properties~\cite{ref:CMB}, but others are tied up in 
degeneracies, insensitivities (e.g.~to the dark energy equation of 
state behavior), and astrophysical foregrounds. Structure growth and 
evolution measures such as cluster counts by the Sunyaev-Zel'dovich 
effect or $X$-ray surveys
and galaxy halo-density and velocity distributions
through 
redshift surveys
need to translate and disentangle observed quantities 
from theoretical ones, through problematic relations such as the 
mass-temperature law and the nonlinear matter power spectrum.  
Asphericities, clumpiness, bias, and foregrounds all play roles.  
Similar difficulties also 
apply to weak
and strong
gravitational lensing, 
Sunyaev-Zel'dovich 
measures of the angular diameter distance,
peculiar velocity 
measurements,
Alcock-Paczy{\'n}ski redshift 
distortions,
etc. Certainly the 
Type Ia supernova method that discovered the acceleration of the 
universe~\cite{ref:SCP,ref:High-Z} is not exempt. 

Type Ia supernova measurements have an advantage in a long track record 
of observations pushing the systematics to lower levels by 
understanding the astrophysical effects, making supplementary 
measurements, and correcting for the intervening quantities, reducing 
them to smaller residual uncertainties~\cite{ref:systSCP}.  The quest for accurate 
cosmological parameter estimation at the percent level requires a well 
designed experiment dedicated to obtaining a systematics-bounded dataset.  
As we seek to pursue the supernova distance-redshift measurements to 
higher redshifts we gain from the increased discrimination between 
model parameters and from degeneracy breaking, but also require more 
stringent understanding and correction of systematics. 

The level of accuracy required goes beyond treatment by 
simple analytic and Fisher matrix methods.  While a rigorous 
treatment of the complexity of astrophysical observations  --from 
detector pixel response to light-curve fitting to non-Gaussian and 
correlated errors-- requires a well crafted end-to-end Monte Carlo 
simulation of the survey, this is perhaps overly detailed and model 
dependent to draw general conclusions.  

In this paper we apply a more illustrative Monte Carlo method to broadly 
investigate the effects on cosmological parameter 
estimation  --both errors and biases-- due to systematic uncertainties 
and biases in the 
supernova magnitudes used in the Hubble diagram, or distance-redshift 
relation.  Section 2 reviews the Hubble diagram.  Section 3 considers 
the effect of an irreducible uncertainty in the form of both constant 
and redshift dependent magnitude systematics, while Section 4 examines 
the effect of a magnitude offset  
bias.  The next three sections trace specific systematics through 
the supernova observations to the Hubble diagram and the cosmology 
fitting: Section 5 
discusses calibration error, Section 6 selection effects
(Malmquist bias), and Section 7 the magnitude de/amplification 
from gravitational lensing.  Section 8 addresses methods for 
countering ``evolution'', or population drift, through comprehensive 
use of spectral and flux time-series data. 

For the Type Ia supernova 
method, this paper presents the basic analysis of the role of systematic errors 
and corrective measures in obtaining accurate as well as precise 
determinations of the cosmological parameters. Any cosmological 
probe or similar supernova survey must carry out such studies in order to 
quote parameter fitting capabilities with rigor.  In the 
end, however, we will have a broad network of results, 
complementary and cross checking and thus synergistically powerful, 
that test the cosmological model and lead us toward understanding 
the fundamental physics responsible for our accelerating universe. 
\section{Tracing Cosmology with Supernovae}\label{sec:sncos} 
Type Ia supernovae have long been recognized as a powerful probe of
cosmological dynamics, particularly in the measurement of its rate of
expansion.  In the SN Ia Hubble diagram, a plot of supernova peak
magnitude versus redshift, the supernova brightness serves as a proxy
for the supernova distance.  The redshift $z=a^{-1}-1$ measures the
scale factor $a$, the size of the universe when the supernova
light was emitted relative to its current size.  
At low redshift, the data provide 
confirmation of a linear relationship between redshift and distance, 
the Hubble law.  The dispersion of the data around the Hubble law
measures how well supernova brightnesses serve as a proxy to distance.
Studies of Type Ia Hubble diagrams have provided convincing evidence
that SNe Ia can serve as standardizable candles, being able to
determine luminosity distances to 5--10\%
\cite{ref:phillips99,ref:Tripp:1999,ref:rpk96}.  Models for SNe Ia must and do
provide a simple theoretical explanation for this observational
homogeneity \cite{ref:hoeflich03}.

When considering distant supernovae, one should recognize that
cosmological distances translate into a lookback time $t$ to when the
supernova explosion occurred.  Thus the distance-redshift measurement
is also a mapping of the cosmic expansion history $a(t)$.  At further 
lookback times (larger distances), secular 
deviations from the linear Hubble law, which represents a constant 
expansion rate, measure the acceleration or deceleration
of the expansion of the universe.  Fitting a set of cosmological
parameters to the data in the Hubble diagram allows precise
discrimination of different cosmological models, together with error
estimation and confidence levels.  This paper uses Monte Carlo
and a flexible
$\chi^2$-based cosmology fitter to simulate data from various
distributions of supernovae, add statistical and systematic errors,
and derive joint probability contours for the cosmological parameters.

The critical ingredients are the background cosmology model 
(\S\ref{sec:cosmod}), the supernova survey characteristics 
(\S\ref{sec:survey}), and the observational and astrophysical 
errors (\S\ref{sec:irrsys}--\S\ref{sec:lensing}).  
Details of the 
fitting procedure are given in \S\ref{sec:fit}. 
\subsection{Cosmology}\label{sec:cosmod}
The luminosity distance to an object at redshift $z$ is given in terms 
of its comoving distance $r(z)$ by $d(z)=(1+z)\,r(z)$.  Astronomers use 
magnitudes, a logarithmic measure of flux, which neglecting astrophysical effects like 
dust absorption,
can be written as 
\begin{eqnarray} 
m(z)&=& 5\log_{10} d(z) \nonumber \\
&+& \left[M + 25 -5\log_{10}\left(H_0/\left(100{\rm km/s/Mpc}\right)\right)\right] \ ,
\end{eqnarray}
where the distance $d(z)$ is made dimensionless by removing the 
Hubble scale $H_0^{-1}$, $M$ is the absolute magnitude of a supernova, 
and the constant in brackets is often notated 
$\cal M$.
The influence of 
the cosmology resides in $d(z)$, or equivalently $r(z)$. 

The comoving distance follows directly from the metric by 
\beq
r(z)=\omk^{-1/2}\sinh\left[\omk^{1/2}\int_0^z dz'/\left[H(z')/H_0\right]\right],
\eeq
where $\omk=1-\Omega_{\rm tot}$, $\Omega_{\rm tot}$ is the total 
dimensionless energy density of the universe,  $\sinh$ is analytically 
continued to $\sin$ for imaginary arguments, and $H(z)$ is the Hubble 
parameter.  Since cosmic microwave background data strongly suggests 
our universe is flat, $\omk=0$, we employ in the rest of this paper 
the appropriate limit, 
\beqa 
r(z)&=&\int_0^z dz'/\left[H(z')/H_0\right] \label{rz} \\ 
&=&\int_0^z dz'\left[\om(1+z')^3 \right.\nonumber \\
& &\ \ \ \ \ \ \  +\ (1-\om)e^{3\int_0^{z'}\frac{dz''}{1+z''}\, 
(1+w(z''))}\left.\right]^{-1/2}.
\label{eq:r(z)}
\eeqa 
Here $\om$ is the dimensionless matter density and $w(z)$ is the equation 
of state, or pressure to energy density ratio, of the other component --  
negative-pressure dark energy. 

Two common parameterizations for the equation of state are $w=w_0+w_1z$ 
and $w=w_0+w_a(1-a)$, where $a$ is the scale factor of the universe. 
The exponential in~(\ref{eq:r(z)}) resolves respectively to 
$(1+z')^{3(1+w_0-w_1)}e^{3w_1z'}$ and $(1+z')^{3(1+w_0+w_a)} 
e^{-3w_az'/(1+z')}$.  In either case we have a three-parameter 
phase space describing the cosmology: $\om$, $w_0$, $w'$ (where the 
time variation $w'$ is taken 
to be either $w_1$ or $w_a/2$; for purely historical reasons, we use $w_1$ in this paper).  
This covers the important quantities 
of the energy density, the dark-energy equation of state (EOS), and a 
measure of the physically revealing EOS time variation.  In addition 
there is the nuisance parameter of the zero offset $\cal M$.  

For a fiducial model we adopt in this paper (except in 
Fig.~\ref{fig:offset}) a model with $\Om=0.3$ in matter and $1-\Om$ 
in a cosmological constant, $w=-1$ (that is, $ w_0=-1$ and $w'=0$).
Section \S\ref{sec:fit} contains a 
discussion of the effect of variation of the model.  The parameters 
are allowed to range freely about the fiducial values.  However, we 
do generally impose on the matter density a Gaussian prior $\sigma(\om)=0.03$, 
reflecting anticipated information from other cosmological probes.
While we believe that this is a reasonable assumption (there are 
recent studies~\cite{ref:ASF} that
claim that a precision around $\sigma(\om)=0.035$ has already been achieved),
we have checked that relaxing this assumption to $\sigma(\om)=0.05$ would roughly increase all
uncertainties on $w'$ given below by less than 50\%,
while leaving the other parameters unchanged (see also~\cite{ref:welal}).
Alternatively, a slightly more
powerful constraint could have been obtained by including as a prior the projected measurement
of the distance to the surface of last scattering by the Planck mission~\cite{ref:fhlt}.
\subsection{Supernova Survey Characterization}\label{sec:survey} 
The observations can be characterized in terms of the distribution in
redshift of the supernova distance data and the dispersion about perfectly 
constant intrinsic peak flux (absolute magnitude) remaining after 
standardization for astrophysical and observational effects.  
For determining cosmological parameters, only 
the relative flux ratio between supernovae at different redshifts, not 
the absolute values, is important, thus the constant offset $\calm$ 
composed of the absolute magnitude and the Hubble constant  
is just a nuisance parameter that needs to be integrated over. 

One promising survey is the proposed Supernova/Acceleration 
Probe (SNAP)~\cite{ref:snap}.  This involves a 2 meter 
telescope in space discovering and following some 2000 
SNe Ia between $z=0.3-1.7$ with an optical/near infrared 
imager and spectrograph.
 
The magnitude dispersion of a given supernova is assumed to be constant and
independent of redshift, for a well designed survey.  We take it to be $\sigma_m=0.15$ for 
all surveys. 
This roughly corresponds to an intrinsic magnitude
dispersion of 0.1 mag (or a 5\% uncertainty in distance) and an
equivalent statistical uncertainty in the determination of the
corrected peak magnitude.
The aggregated statistical error can be reduced by increasing the 
survey size to boost the number of supernovae.  Note that at redshifts 
$z\la0.8$ SNAP essentially 
follows all supernovae in the volume, so for these redshifts the 
sky area or survey lifetime would need to increase to gain statistics. 

Since the sensitivity of the data to the cosmological parameters depends 
on redshift (for example at low redshift the distance reduces to $d=z$  
for all cosmological models), the redshift distribution of the supernovae 
is important.  The major effect is from the survey depth, $z_{max}$; 
optimization studies show the parameter estimations to be fairly 
insensitive to the exact distribution so long as the full 
redshift range is covered \cite{ref:huttur,ref:fhlt}.  Thus other 
observational and instrumental considerations can be taken into 
account for the distribution without harm to the science results. 
\begin{table*}
\begin{minipage}{155mm}
\caption[]{The redshift distribution $N(z)$ of the 2000 SNe employed 
from a fiducial SNAP survey. 
The redshifts $z$ given in the table
correspond to the upper edges of each bin.}
\label{tab:z}
\begin{center}
\begin{tabular}{|c||r|r|r|r|r|r|r|r|r|r|r|r|r|r|r|r|r|}
\hline
$z$ & 0.1 & 0.2 & 0.3 & 0.4 & 0.5 & 0.6 & 0.7 & 0.8 & 0.9 & 1.0 & 1.1 & 1.2 
 & 1.3 & 1.4 & 1.5 & 1.6 & 1.7 \\
\hline
$N(z)$ & 0 & 35 & 64 & 95 & 124 & 150 & 171 & 183 & 179 & 170 & 155 & 142 & 130 
& 119 & 107 & 94 & 80\\
\hline
\end{tabular}
\end{center}
\end{minipage}
\end{table*}

For the space-based mission  
we adopt a fiducial distribution shown in Table~\ref{tab:z}, which we will call the SNAP
distribution in the following. 
When considering shallower surveys we 
cut and rescale the SNAP distribution: truncating 
the SNAP distribution at the new $z_{max}$ and then multiplying $N(z)$
by the factor needed to keep the total number unchanged from 
the original, e.g.~2000 supernovae.  This ensures that we can compare 
surveys based on their redshift reach, with their statistics on an equal 
footing.  As we discuss in \S\ref{sec:irrflat}, 
the parameter estimation precision 
is not driven by statistics, i.e.~numbers of supernovae, so we consider the 
total number fixed at 2000, well within the capabilities of SNAP.  
We have also checked that adopting instead a form $N(z)$ of some 
proposed ground-based survey affects the results by less than 
a few percent. 

Additionally, in all cases we include a very low redshift (``local Hubble flow'') 
group of supernovae, 300 between $z=0.03-0.08$, such as will soon 
be available from the Nearby Supernova Factory \cite{ref:snf}.  These prove 
important for marginalizing (averaging) over the extra parameter $\calm$ 
to reduce the parameter phase space.  
\subsection{Cosmological Parameter Fitting}
\label{sec:fit}
Given the elements of the previous two subsections, we can generate 
Monte Carlo realizations of supernova magnitude data vs.~redshift, 
i.e.~Hubble diagrams.  These are then fit with the set of 
cosmological parameters using an unbinned $\chi^2$ minimization method.  Two 
independent codes, one using the Minuit minimization package from the 
CERN library~\cite{ref:minuit} and 
the other a CERN adaptation of a NAG routine~\cite{ref:NAG}, 
have been checked against each other. 
Each generates the best fit to the data within the four dimensional 
parameter space $\{\Om,w_0,w',{\calm}\}$ and the contours of 68\% 
(or whatever level) confidence.  The two-dimensional plots shown are marginalized over the other 
two parameters. In general the errors 
are non-Gaussian and asymmetric.  Where numbers are quoted as $1\sigma$ errors, they
refer to the 68\% confidence level parabolic error on that parameter, 
marginalizing over the rest of the likelihood space. 
 
Note that the error contours depend not only on the data errors 
but also on the background, fiducial cosmology.  This is discussed and 
illustrated in \S\ref{sec:offset}.  So in fact there is no 
unique parameter estimation precision associated with a given 
survey, even if the data properties, systematics, and priors 
are all specified.  The numbers we quote are for $\Om=0.3$, a flat universe,
and the dark energy being a cosmological constant, unless stated otherwise.  

Errors on dark-energy properties go down as its energy density, and hence 
effect on the expansion, increases.  For 
example, if $\Om=0.25$ the SNAP estimates of $w_0$ and $w'$ improve 
by 13\% and 8\% respectively with respect to the fiducial case
$\Omega_m=0.3$.  Making $w(z)$ more positive than 
the fiducial value $-1$ over 
the redshift range of the data, either by increasing $w_0$ or 
taking a positive $w'$, raises the dark energy density at those 
redshifts.  So this also increases the sensitivity of parameter 
estimation, and hence precision.  Thus, if the effective equation-of-state 
parameter is less
negative than $-1$, the cosmological model 
we have adopted gives a conservative  
assessment of the supernova method as a cosmological probe. 
\section{Uncorrelated Systematic Uncertainties}
\label{sec:irrsys}
In addition to the statistical errors already discussed, systematic 
uncertainties need to be taken into account.  These arise from 
imperfect sources (e.g.~population evolution), imperfect detectors 
(e.g.~calibration 
errors), and intervening astrophysics (e.g.~dust, gravitational lensing). 
Detailed discussion of methods developed for bounding these through precise and 
multiwavelength observations and like-to-like subsample comparison is 
beyond the scope of this paper (see \S\S\ref{sec:calib}--\ref{sec:evolution} for an 
introductory treatment).  
Rigorously, this requires a 
comprehensive Monte 
Carlo simulation with a multitude of model dependent parameters 
characterizing the instruments, astrophysics, survey strategy, etc. 
Here we concentrate on principles derived from a more general analysis 
of illustrative systematic error behaviors. 

To draw conclusions about the impact on cosmological parameter 
determination within a given survey, we investigate two general forms 
of systematics.  The first is 
a random dispersion that is irreducible below some magnitude error over a 
finite redshift bin.  We adopt a bin width $\Delta z=0.1$ as a rough 
estimate of the correlation of cosmic conditions and instrumentation.  The second 
systematic is a (possibly redshift dependent) magnitude offset that acts 
coherently on all supernovae (discussed in \S\ref{sec:offset}).  

The size of these systematic uncertainties will depend on details 
of the survey depth and strategy and the instrument suite.  The SNAP 
mission is specifically designed in these details to bound the sum 
of all known and proposed systematics below 0.02 mag.  Besides this 
fiducial we also consider the effect of larger errors over the same 
redshift range and the case of a shallower survey such as could be 
achieved from the ground.  Frieman et al.~\cite{ref:fhlt} and Linder \& 
Huterer \cite{ref:linhut} discussed the generic need for observations 
to reach beyond $z\approx1.5$ to detect the time variation $w'$, 
improve precision on $w_0$, and, most relevant to this paper, 
immunize against systematics. 
\subsection{Random Irreducible Systematic -- Flat} 
\label{sec:irrflat}
An error that is inherent 
to the measurement process of a supernova would not be statistically 
reduced with greater numbers of measurements.  Hence we refer to it 
as irreducible. 
Examples of this class could be calibration errors, and errors in galaxy
subtraction coming from the lack of perfect knowledge of the point spread 
function.

To simulate such an error, we introduce an irreducible magnitude 
error, $dm$, on the measurement in each supernova redshift bin.  
In essence, this models an error in such a binned approach for 
each filter type on the camera whose peak response is roughly spaced in redshift 
by $\Delta z $ = 0.1.  The error is added in quadrature to the canonical 0.15 
mag intrinsic magnitude dispersion per SN:

\begin{equation}
\sigma_{m} = \sqrt{\frac{0.15^{2}}{N_{bin}} + (dm)^{2}}\ ,
\end{equation} 
where $N_{bin}$ is the number of supernovae in a 0.1 redshift bin (see, e.g., Table~\ref{tab:z}).

\begin{figure}
\includegraphics[width=84mm]{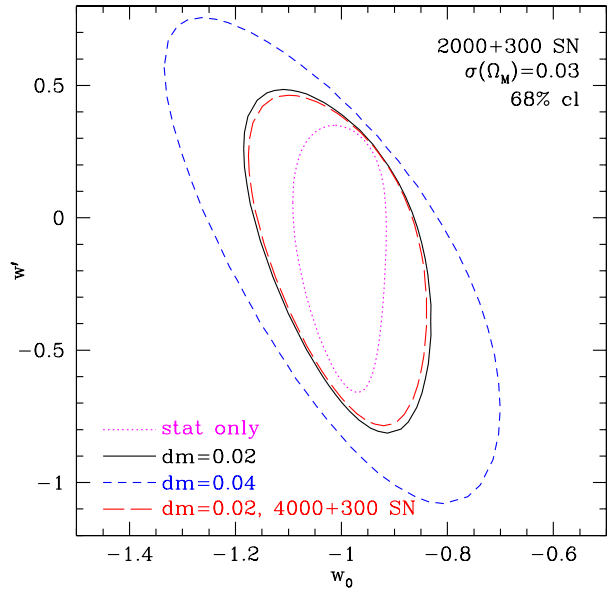}
\caption{Dark energy parameter contours for three irreducible systematic cases. 
The central reference contour has only the intrinsic statistical error. Note the 
substantial increase in parameter error as the systematic is increased from 
$dm$ = 0.02 to 0.04.  The two closely paired contours show that doubling the 
number of SNe will only increase parameter accuracy by $\approx$ 5$\%$.}
\label{fig:irrflat.ww1}
\end{figure}

Figure \ref{fig:irrflat.ww1} shows the 68$\%$ joint probability contours 
in $w_0$ and $w'$
for SNAP's distribution of 2000 (extending up to $z=1.7$) plus 300 nearby supernovae
under irreducible magnitude errors 
$dm=0.02,0.04$.  
To show the dominance of the systematic contribution 
over the statistical error we also double the number of supernovae 
in each bin (except the first bin representing the SNfactory sample). 
A fiducial input cosmology of 
$\om = 0.3$, $w_{0} = -1.0$, $w' = 0$ is used along with a prior on 
$\om$ of 0.03. The 
intrinsic statistical error contours are also plotted for reference. 

Let us note three important points about the systematic uncertainties 
evident from the figure: 1) they are the 
dominant source of error, 2) they impact parameter determination in 
nontrivial ways, and 3) allowing systematics to exceed 0.02 
mag can strongly affect parameter estimation. 

Overall one clearly sees that observing thousands more 
supernovae will not help in cosmological parameter estimation.  Only 
by tighter bounds on the systematics can one improve precision.  The 
systematics impose an error floor and become the dominant contribution 
over the statistical error for 
\beq 
N_{\rm bin}>(0.15/dm)^2. \label{nbin} 
\eeq
For $dm=0.02$ mag this works out to about 55 supernovae per bin.  It is 
worth noting some reasons, however, why one might want to exceed this 
number somewhat.  These include: 

\begin{enumerate} 
\item Subsamples, 
e.g.~collections of supernovae with similar spectral properties, 
allow carrying 
out a like-to-like comparison to identify and bound systematic 
uncertainties.  Examination of supernovae within a bin, i.e.~all at the 
same redshift, probes systematics while analyzing like subsamples at 
different redshifts probes cosmology more cleanly. 
\item Lowering statistical errors well below the systematics floor 
reduces the total, quadratic sum error.  For example twice as 
many supernovae as the break even number given by eq.~(\ref{nbin}) 
lowers the total error to within 22\% of the floor. 
\item Additional data can mitigate the effects of mis-estimation 
of the coherence scale, i.e.~effective bin size, of the systematic. 
That is, since a systematic of 0.02 mag over a bin size of 
$\Delta z=0.05$ is roughly the same as $dm=0.014$~mag over $\Delta z=0.1$, 
more data ensures that statistical errors will not 
if the systematic error is smaller than 0.02 mag.
\item The complete supernova sample will include events that suffer
significant host-galaxy dust absorption and gravitational
demagnification.  Beyond our systematic concerns, the light-curve and
spectral measurements of these objects will carry reduced statistical
weight due to their fainter appearance.  An increased number of supernovae
per redshift bin will make up for the expected range of data quality.
\end{enumerate} 
Overall, a contingency of roughly a factor of two more supernovae 
should prove satisfactory. 

For the second point, Fig.~\ref{fig:irrflat.ww1} shows how a 
constant irreducible systematic of 0.02 magnitudes increases the 
parameter estimation errors.  Note that the systematic does not 
simply scale up the contour but rather stretches it along the 
parameters' correlation or degeneracy direction (very roughly defined 
by $w'+4w_0$=const \cite{ref:huttur,ref:welal,ref:linhut}).  Relative 
to the pure statistical error case, a 0.02 mag systematic increases 
the uncertainty in $w_{0}$ by a factor of 2 and $w_{1}$ by $26\%$. 
The effect of systematics on parameter estimation can also be a 
strong function of depth of the survey and underlying cosmological 
model, as discussed in the following sections. 

Increasing the magnitude error beyond 0.02 mag, to 0.04, strongly 
degrades the precision with which the dark-energy parameters can be 
recovered.  The constraint of $w_0$ suffers an additional 78\% and 
$w'$ incurs an extra 44\% increase. Both of them become so imprecise 
that the limits are not useful. 
Therefore, bounding the systematics to 0.02 magnitudes is an 
important science goal for an experiment that hopes to detect the 
time variation of the dark-energy equation of state.  Instruments 
and observation strategies must be specifically designed with this 
in mind. Garnering a sufficiently rich, well calibrated, and 
homogeneous set of data allows control for astrophysical and 
measurement effects so as to leave behind only a less than 0.02 mag 
residual. 

\begin{figure}
\includegraphics[width=84mm]{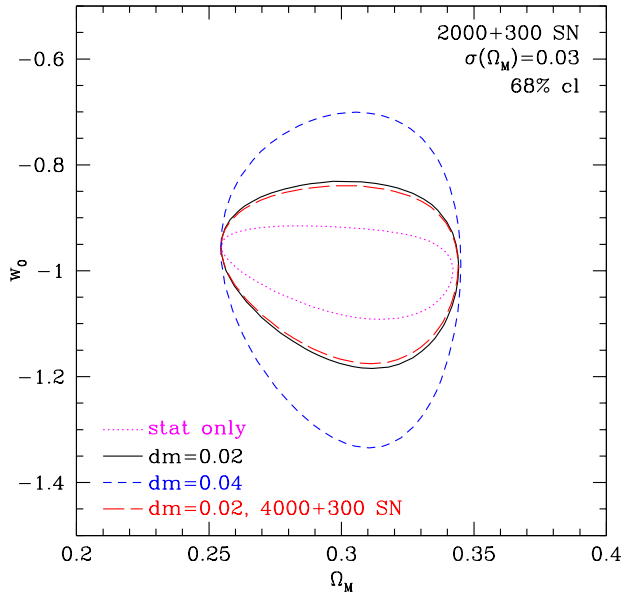}
\caption{Contours for the mass density $\Omega_{M}$ versus the present 
dark energy equation of state 
$w_{0}$ for three irreducible systematics. As the systematic increases, 
the large increase in error 
for the equation of state is clearly apparent. 
The $\Omega_{M}$ parameter error stays constant since a prior constraint of 
$\sigma(\Omega_{M})$ = 0.03 dominates the estimation.}
\label{fig:irrflat.omw}
\end{figure}

Figure \ref{fig:irrflat.omw} illustrates the same systematics in the 
$w_{0} - \Omega_{M}$ plane.  
Again, the statistical errors are only 
a minor contribution to the total. Because the parameter estimation 
of $\Omega_{M}$ is dominated by the imposed prior 
$\sigma(\Omega_M)=0.03$, 
the systematic acts in the $w_0$ direction.  The uncertainty in 
$w_{0}$ increases over the purely statistical error by a factor of 2 
and 4 when $dm$ = 0.02 and 0.04, respectively.  Note that no prior 
assumption about $w'$ was imposed, so the equation of state variable 
is $w_0$ not merely a constant $w$.  Given that we are seeking to 
test the nature of dark energy, not assume it, and that there is 
no compelling theoretical model that predicts a constant value 
(other than possibly $-1$), we do not believe that results involving 
a constant $w$ serve any useful purpose for precision cosmology.
\subsection{Random Irreducible Systematic -- Linear in Redshift} 
\label{sec:irrlin}
Observations extending to the redshift depth necessary to probe 
dark energy, $z\ga1.5$, are challenging and one might well expect 
some possible errors to be exacerbated with increasing redshift. 
The restframe optical emission of supernovae shifts to the observer 
frame as $1+z$, with key spectral features at $z=1.7$ 
approaching 1.7 microns, so infrared capabilities are crucial to 
these high redshift observations.  SNAP is specifically designed 
to include high-precision photometry out to 1.7 microns, but 
residual uncertainties enter.  
For example, models 
for Hubble Space Telescope (HST) spectrophotometric standards can 
disagree up to 1\% at 1.7 microns~\cite{ref:bohlin02} due to 
the lack of precision exo-atmospheric, spectrophotometric measurements 
in the near infrared (NIR).  

Since wavelength maps to redshift, larger
errors in the infrared translate 
to increasing uncertainties at higher redshifts. To simulate the 
effect of such uncertainties on
parameter determination, a linearly increasing systematic, 
$dm = \delta m\left(z/z_{max}\right)$, 
is adopted. As before, this is added in quadrature to the 
intrinsic 0.15 mag statistical error for supernova peak magnitudes in a binned 
approach: 
\begin{equation}
\sigma_{m} = \sqrt{\frac{0.15^{2}}{N_{bin}} + \left(\delta m
\frac{z}{z_{max}}\right)^{2}}.
\end{equation}
Parameterization in terms of $z_{max}$ and $\delta m$ allows us to 
perform trade studies on other surveys with different redshift depths 
and infrared error limits.  The SNAP design bounds the uncertainty 
for all relative photometric measurements to 0.02 mag, so the systematic 
ramps up as $dm=0.02(z/1.7)$. 

\begin{figure}
\includegraphics[width=84mm]{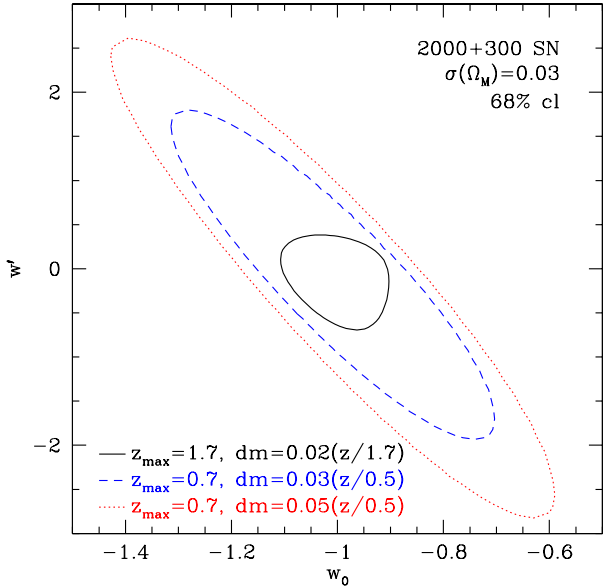}
\caption{Dark-energy parameter contours with a linear systematic that increases with redshift.  
The SNAP contour includes a systematic of $dm = 0.02(z/1.7)$ benefiting from the 
experiment's high
precision photometry and survey depth. Also plotted are two systematics simulating ground-based 
experiments that have larger photometric errors into the infrared and an effective 
redshift depth limited 
by the atmosphere. 
}
\label{fig:irrlin.ww1}
\end{figure}

Figure~\ref{fig:irrlin.ww1} shows parameter estimation confidence 
contours for some different survey depths and amplitudes of the linear 
systematic. 
These include SNAP, modeled as a space-based experiment with 
$z_{max}$ = 1.7 and $\delta m$ = 0.02 mag, and a ground-based experiment with 
an atmosphere-limited effective redshift of $z_{max}$ = 0.7 and either an optimistic 
$\delta m$ = 0.04 mag (corresponding to the line with $dm=0.03*(z/0.5)$ 
in Fig.~\ref{fig:irrlin.ww1}) or a less challenging 
0.07 mag (corresponding to $dm=0.05*(z/0.5)$ 
in Fig.~\ref{fig:irrlin.ww1})~\footnote{These two values
bracket the precision claimed for the proposed Essence supernova survey~\cite{ref:ess}.}.
The surveys were 
otherwise identical, with the same number of supernovae (the SNAP distribution 
of Table \ref{tab:z} was cut at $z_{max}$ and rescaled to total 2000 
supernovae) and prior on the matter density. 

The figure dramatically 
demonstrates that a deeper survey in redshift with the infrared 
capabilities to control systematics (especially dust extinction), opened 
up by the move to space, provides a superior lever arm in 
determining the cosmological parameters.  Ground-based surveys, however, are 
limited for obtaining a homogeneous, complete data set to $z_{max}\la0.7$ 
(see \S\ref{sec:malm}); increased statistics, such as 
provided here with 2000 supernovae, will not help.  
Figure~\ref{fig:irrlin.ww1} shows that even in the optimistic ground-based case,
the parameter uncertainty increases by a factor of 3 in $w_{0}$ and a factor 
of 4 in $w_{1}$.  The other case, $\delta m=0.05$, perhaps more 
realistically simulates the increase in uncertainties due to 
atmospheric absorption and night sky emission as one ventures into the 
infrared from the 
ground.  Here the parameter estimation degrades by 24\% in $w_{0}$ and 
36\% in $w_1$, just relative to the optimistic ground-based survey.

\begin{figure}
\includegraphics[width=84mm]{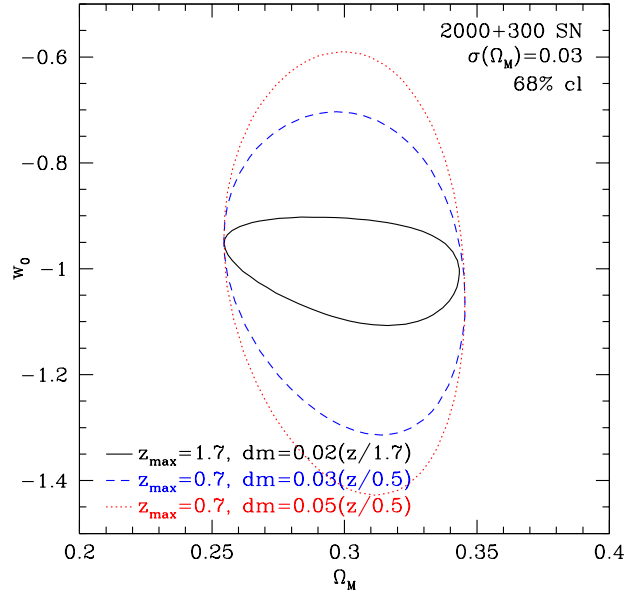}
\caption{Error contours in the $\Omega_{M} - w_{0}$ plane for systematic errors 
linearly increasing with redshift. The SNAP measurement of $w_{0}$ shows the 
advantage of a deeper space-based survey with better photometric accuracy. 
There is a negligible change in the $\Omega_{M}$ measurement since the 
parameter is bounded by prior information.
} 
\label{fig:irrlin.omw}
\end{figure}

The equivalent blowing up of the confidence contours in the 
$w_0 - \Omega_{M}$ plane is shown in Fig.~\ref{fig:irrlin.omw}.  
Clearly, redshift depth and dedicated instrument design to bound  
systematic uncertainties are crucial to achieving precision cosmological 
parameter determination.  

While the two models for systematics discussed 
in the last two subsections are not exhaustive, they do give a broad 
feel for different, physically motivated behaviors.  These analyses 
indicate that the survey requirement of bounding systematics below 0.02 mag 
is both necessary and sufficient for the science goals of constraining 
the dark energy equation of state and its possible variation. 
\section{Systematic Biases}\label{sec:offset}
The other category of systematic effects considered is a coherent shift in 
magnitude for all supernovae at a given redshift. This could arise from 
broad astrophysical or detector issues such as residual uncertainties 
in intergalactic absorption, or selection effects such as Malmquist bias 
(see \S\ref{sec:malm}).

The effect of an offset in magnitude is a bias in the best fit parameters. 
That is, the data no longer guide the observer to the true underlying 
cosmology but rather to a false model.  Obviously this is of great 
concern since we seek not only precise but accurate answers.  
Interestingly, the bias can actually somewhat reduce the dispersion 
around the best (incorrect) fit, 
since as mentioned in \S\ref{sec:fit} the errors depend on 
the location within the parameter space.  Thus one could be put into the 
situation of finding a wrong answer very precisely.  

Bounding offsets below 0.02 mag, however, ensures that bias of the 
fit parameter from the true value is negligible, i.e.~less than half 
the random, dispersive error.  Furthermore, one is fortunate in that 
it is not the entire amplitude of the offset that alters the 
cosmological parameters of interest, but rather only its variation 
from the mean over the survey depth.  Thus a constant offset has no 
effect on the cosmological and dark energy parameter estimation,
and a linear increase to 0.02 mag at the maximum redshift is equivalent
to a $-0.01$ mag shift at low redshift and a $+0.01$ mag 
shift at high redshifts.  
This point is sufficiently important that 
we present a formal proof in the Appendix.
\subsection{Monte Carlo Analysis of Systematic Magnitude Offsets}
Upon adding a constant 0.02 mag offset to all
supernova magnitudes 
we verified by Monte Carlo that the best fit value of $\mathcal{M}$ was biased from an 
input of 0 to 0.02 while the fits for the cosmological parameters returned 
the true values and had errors indistinguishable from the case without 
such a systematic bias.

\begin{figure}
\includegraphics[width=84mm]{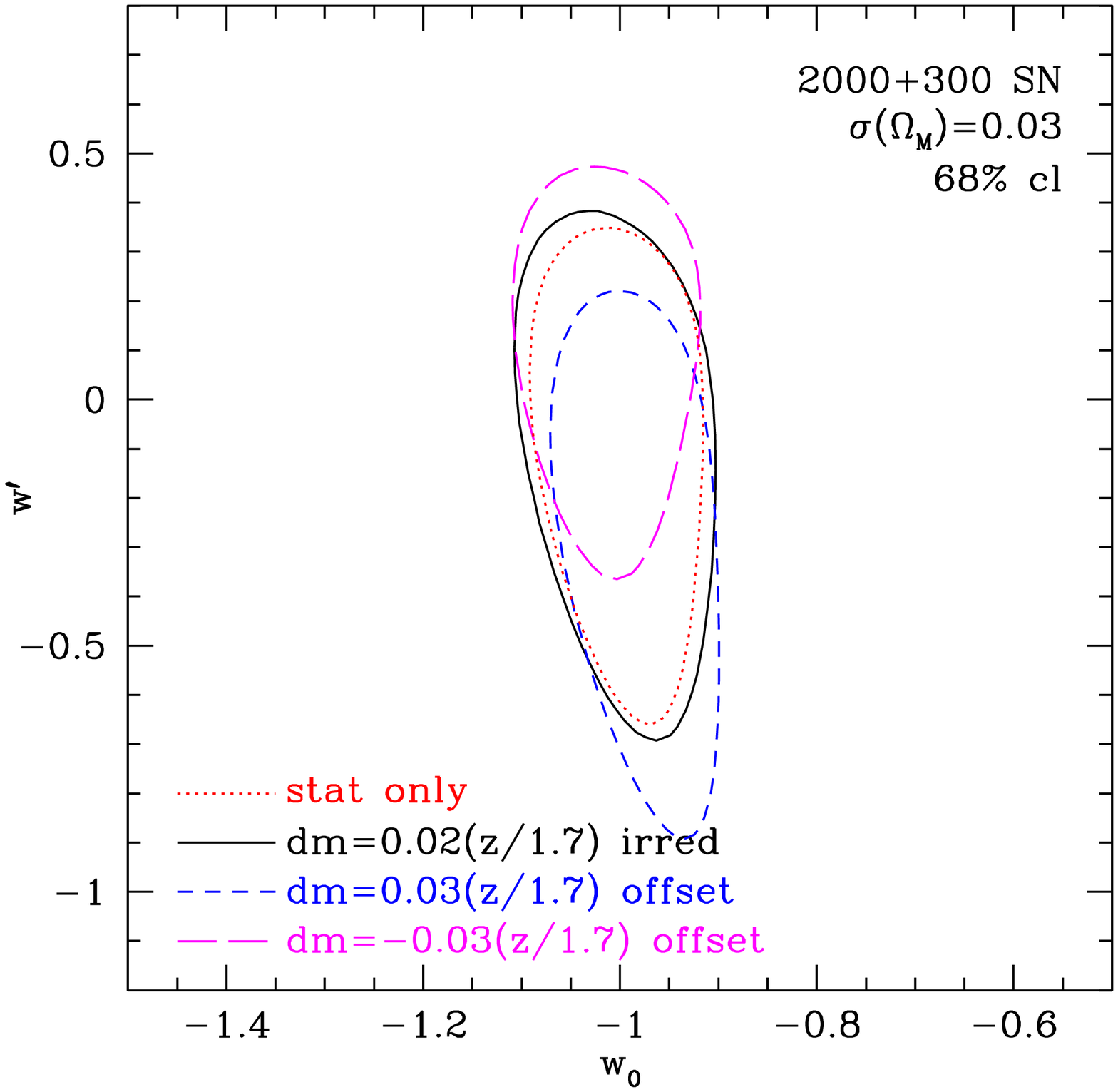} 
\includegraphics[width=84mm]{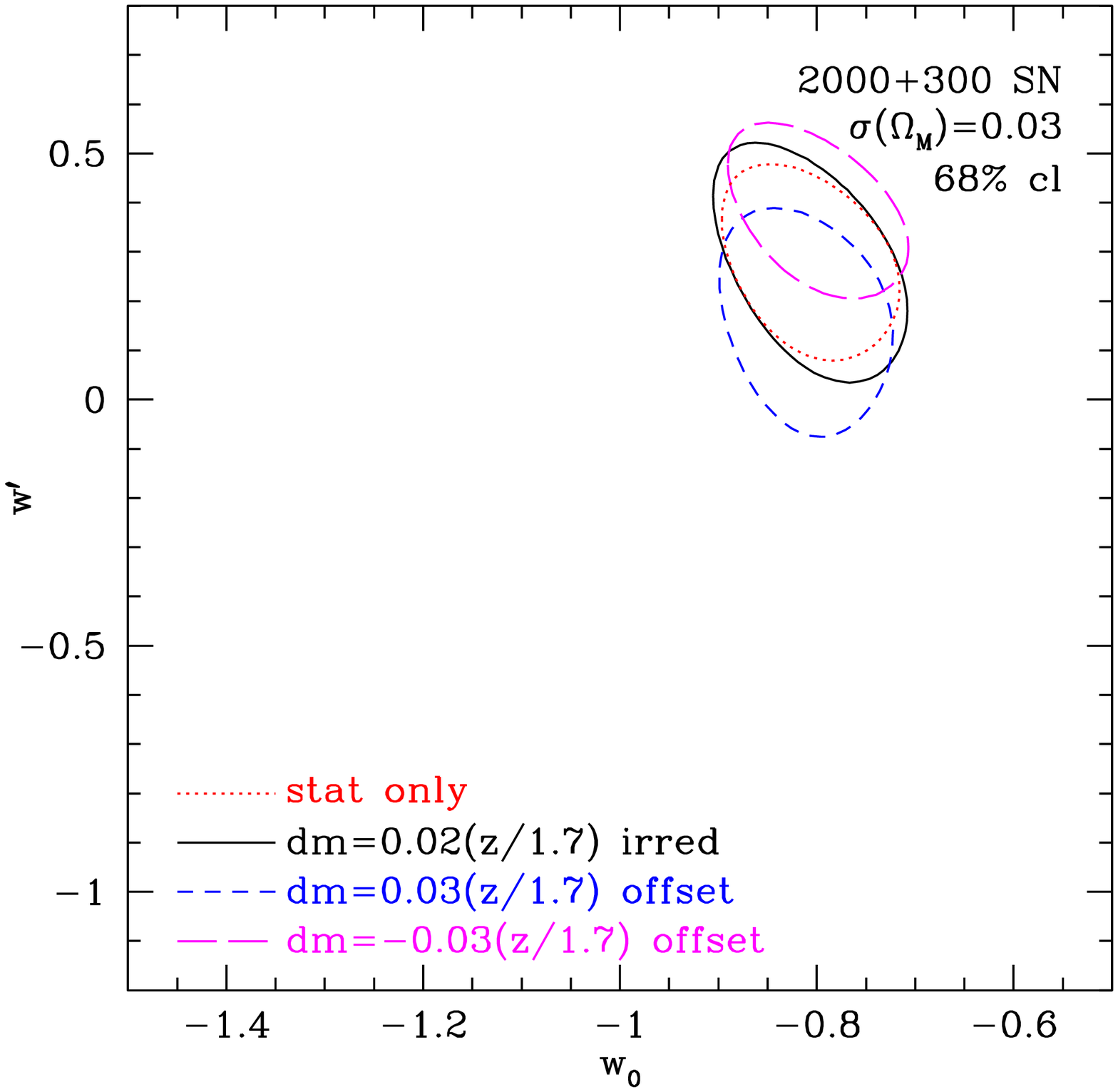}
\caption[kk]{Dark-energy parameter contours after a magnitude bias is introduced. The 
solid lines include systematic errors but no offset, while the dashed curves
have an offset but no systematic uncertainties. The upper plot corresponds
to the fiducial universe with $w_0=-1$ and $w_1=0$, while the lower plot
corresponds to the SUGRA-inspired scenario~\cite{ref:braxm,ref:welal} 
with $w_0=-0.8$ and
$w_1=0.3$. Note the large dependence of the precision attainable on the underlying
cosmology.}
\label{fig:offset}
\end{figure}

To investigate bias in the parameters and its amplitude relative to its 
random error, we introduced a magnitude offset that 
linearly increases with redshift.  Specifically, we adopt $dm=\pm0.03(z/1.7)$.  
Figure \ref{fig:offset} shows the results for two different underlying 
cosmologies.  
The upper plot has an input model where the dark energy 
is a cosmological constant: $w_0=-1$, $w_1=0$, while the lower plot illustrates 
a model with time-varying equation of state for the dark energy: $w_0=-0.8$, 
$w_1=0.3$, corresponding to a supergravity inspired model \cite{ref:braxm,ref:welal}.

In each case the dotted contours give the parameter estimation assuming only 
statistical errors on the SNAP supernovae.  One can clearly see the dependence 
of errors and correlations on the location in parameter space, as discussed in 
\S\ref{sec:fit}.
The effect of a random irreducible error, such as from \S\ref{sec:irrlin}, 
is to stretch the confidence contours, generally unequally for the two 
parameters. Such an expansion, shown by the solid contours, gives the increase 
in dispersion due to the extra random uncertainty. 

However the effect of a coherent offset in magnitude, shown by the dashed 
contours, is very different.  Here the contours are shifted in the 
$w_{0} - w_{1}$ plane in the presence of the bias, with the positive magnitude bias 
shifting the contours negatively in $w_{1}$.  Little bias is seen in the $w_0$ 
direction because the supernova 
magnitude offset is highly correlated with the time variation $w_1$ and not so much 
with the single, present value of the equation of state.  This can be seen 
as follows: Increasing the magnitude is 
dimming the supernovae; this means they have a greater luminosity distance at a given redshift
than allowed by the fiducial cosmology,
corresponding to more expansion of the universe and 
hence a more potent acceleration, provided by a more negative equation of 
state.  Because the magnitude offset is here increasing with redshift, one 
requires a time varying change in the equation of state, i.e.~a more 
pronounced $w_1$.  (Recall a magnitude offset that is constant with 
redshift can be absorbed wholly into the $\calm$ parameter, while a 
different $w_0$ does not lead to a monotonic magnitude offset.) 

Also note that the biased 
confidence contours alter their shape and size slightly with respect to the 
unbiased, purely statistical error case.  As discussed in \S\ref{sec:fit}, 
this is the result of the different location, and hence sensitivity, 
in parameter space.  In fact, the linear bias may even decrease the 
statistical uncertainty.  But roughly one can consider the effect of 
an offset magnitude systematic as shifting the statistical error 
contour to a biased best-fit value in the parameter space. 

To guard against biased parameter determination one needs to bound systematic 
effects that give rise to redshift dependent 
magnitude offsets.   In general, improved accounting for systematics by 
extending observations to high redshift and into the infrared, requiring a 
space based instrument, constrains such errors.  
Taking this into account, we consider linearly rising offsets reaching 0.02 and 0.04~mag at
$z_{maz}=1.7$, and 0.04 and 0.07 at $z_{max}=0.7$. 
The first one
represents the SNAP systematics goal and 0.04~mag illustrates the efects of
failing to achieve it.
The second 
two, ground, offsets are larger in magnitude to 
simulate the inherent difficulties in precision measurements through the 
atmosphere of higher redshift supernovae whose emission is shifted into 
the near infrared.  The corresponding supernova distribution is taken to 
extend out to $z_{max}=0.7$ where ground based surveys run into degraded light 
curves and Malmquist bias due to the atmospheric limitations. The size of 
the ground offsets are chosen to give residual uncertainties slightly better and worse than
proposed for the Essence supernova survey \cite{ref:ess}.

As discussed before, the strongest effect is on the time variation of 
the equation of state, $w_1$. The four cases respectively give biases of 
$0.37\sigma$, $0.67\sigma$, $0.75\sigma$, $1.2\sigma$, where $\sigma$ is
the corresponding statistical error for each case.
How much bias is acceptable is somewhat subjective.  If we require that 
the bias should not exceed $0.5\sigma$, then we see that even the ambitious 
ground-based survey fails this for $w_1$, and we cannot tolerate 0.04 mag 
offset for the deep, space-based survey.  But a random irreducible systematic 
of these same amplitudes is far more damaging to parameter estimation than 
the coherent offsets discussed in this section.  As mentioned in 
\S\S\ref{sec:irrflat},\ref{sec:irrlin}, the depth and wavelength coverage 
accessible to a space-based survey substantially immunizes it against such 
uncertainties.
For example the estimation of $w_1$ degrades by a factor of 
four when $z_{max}=0.7$ but only by 6\% when $z_{max}=1.7$. 
\subsection{Low-Redshift Supernova Calibration Offset}
A discontinuity in calibration between the proper SNAP sample and 
the low redshift Nearby Supernova Factory (SNF) measurements
would present a different type of offset error.  
The SNF data play a critical role in reducing the uncertainty 
on $\calm$, and hence $w'$ through the correlation of these 
parameters.  Since we conjoin datasets from two separate 
projects it behooves us to investigate a magnitude offset caused 
by differences in the observing methods and equipment.

To simulate such an effect we introduce a constant offset 
to all SNF supernovae relative to the SNAP dataset.  Because the
offset is at a single, low redshift, the major effect is on $w_0$.
For an offset in the range 0.01--0.05 mag, the bias in fitting the 
$w_0$ parameter amounts to a fraction $1.2(dm/0.02)$ of the 
statistical error.  So to keep the bias less than the statistical 
error, the offset should be restricted below 0.02 mag.  
Experimentally, spectrophotometric observations of the same standard stars by both 
SNF and SNAP should limit the offset below 0.01 mag.  We have found that
the biases in the other parameters, $\om$ and $w'$, are indeed much 
smaller as a fraction of their statistical errors. 

\section{Calibration Uncertainty}
\label{sec:calib}
The calibration procedure for a survey is an important 
potential source of systematic uncertainties.  Since calibration 
enters so early in the data pipeline, these uncertainties can 
propagate through several stages. 
For instance, consider a calibration uncertainty on 
the blackbody temperatures, and hence fluxes, of two or more calibration sources 
and their correlations. That error would affect the
measured fluxes of supernovae directly, 
as well as indirectly through the K-correction and the 
extinction corrections.
The subsequent magnitude errors then affect the cosmological 
parameter determination.

Due to the complexity of the problem, we go beyond our previous 
analytic models and develop a complete Monte Carlo simulation.  
The central ingredient is a calibration model (like the two 
blackbody model just mentioned) with a certain number of
free parameters (in this case, the temperatures of the 
blackbodies), their uncertainties, and the correlations among
them. Realizations of calibration parameters are fed to 
a simulation of a SNAP-like mission. This includes statistical 
observational errors, multiband measurements, and the 
appropriate treatment of reddening and K-corrections.
The result of each pass of the 
simulation is a set of SNe with their
measured magnitudes and errors; many realizations then 
provide the individual supernova magnitude errors and 
correlations.  The cosmology fitter uses this error 
covariance matrix to generate uncertainties and covariance 
contours in the cosmological parameters.  
\subsection{The Monte Carlo Simulation}
In the Monte Carlo simulation, the supernova redshift 
distribution and cosmological model are as given in 
\S\ref{sec:sncos}.  As before, 
only the component of the calibration error which varies 
with redshift or, equivalently, with wavelength, matters 
for cosmological parameter estimation.

Differential dimming of the supernova magnitudes in the various
observed wavelengths by dust is simulated using the standard 
parameterization of Cardelli, Clayton and Mathis 
\cite{ref:cardelli}. Values for the extinction value, 
$A_V$, come from a distribution obtained from a Monte Carlo
simulation~\cite{ref:commins} that places SNe in random 
positions of a galaxy with random orientation with respect to
the line of sight. Values for the global extinction coefficient, 
$R_V$, are drawn from a Gaussian distribution centered at 3.1 
and with standard deviation 0.3. 

Once a set of parameters defining the calibration is chosen, 
it is used to compute zero points for each filter, which in turn
enter the cross-filter
K-corrections~\cite{ref:KGP}. The standard SNAP filter set has 
been simulated, although simplified to square filters.  These 
involve nine filters logarithmically spaced and 
broadened in $1+z$~\cite{ref:omnibus}. 
For each SN there is significant data in a minimum of three and a maximum 
of nine filters, depending on redshift.
The flux in each filter is further smeared with an uncorrelated 
2\% error to account for statistical errors.

All optical and near-IR (in the SN frame) photometric data for 
a given SN are used to fit for two parameters: its magnitude 
and its extinction parameter $A_V$.  
By contrast, $R_V$ is assumed in the fit to be constant at 3.1. 
It has been found that this assumption introduces only a small 
bias in $w'$ in the final result without affecting the result for 
$w_0$. As an example, sampling $R_V$ from a Gaussian distribution 
centered at 3.5 and with standard deviation 0.3 in the generation 
stage, while keeping $R_V$ fixed at 3.1 in the fitting stage, 
results in biases of $-0.15$ in $w'$ and $-0.002$ in $w_0$.

After many realizations of the calibration parameters, enough 
statistics are collected to determine the central value and 
variance of the magnitude of each SN, as well as the correlations 
among them. These are input to the cosmology fit, along with an 
error of 0.15 magnitude added in quadrature to each supernova. 
This accounts for the statistical errors and the natural dispersion 
of the SN intrinsic magnitude.  As previously, a flat $\om=0.3$ 
plus cosmological constant universe is fiducial, along with 
a Gaussian prior of 0.03 on $\om$. 
\subsection{Calibration Models}
Two calibration models have been studied, broadly describing 
the classes of calibration procedures.  In the first one 
\cite{ref:lampton}, a single calibration source is taken as 
reference for all wavelengths, in the optical and infrared 
(IR) regions. The source is parameterized as a blackbody with 
a temperature $T$, known with a precision $\Delta T$. This 
model can be thought of as representing a hot white dwarf whose 
spectrum is well understood in both the optical and IR regimes. 
Typical values for $T$ would be around 20000~K, with 
uncertainties in the few percent range.

In a second, possibly more realistic, model \cite{ref:susana}, 
two calibrators are employed: one with temperature 
$T_1\pm\Delta T_1$ for the optical region, $\lambda \leq 1 \mu{\mathrm m}$, 
and another with $T_2\pm\Delta T_2$ for the infrared 
region, $1\mu{\mathrm m} < \lambda < 1.7\mu{\mathrm m}$. The errors in the two temperatures
are taken to be correlated with correlation coefficient $\rho$. 
The optical calibration could be from a
hot white dwarf, while the near infrared calibration could 
correspond to a NIST standard or a solar equivalent.
In both cases, $T_2$ would be a few thousand Kelvin. The errors 
would again be a few percent. Although the two calibration 
sources are not expected to be correlated in themselves, 
correlations can enter either from common instrumental systems 
and data reduction or from the process of
connecting the optical and near-IR calibrations.
\subsection{Results}
\begin{figure}
\includegraphics[width=84mm]{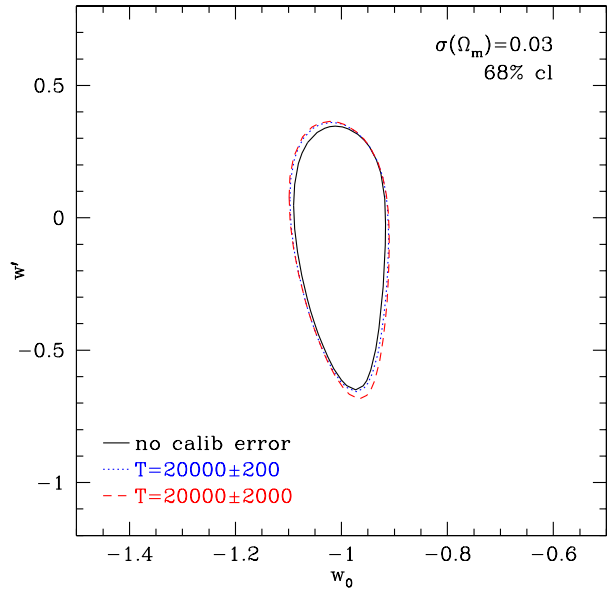}
\caption[]{Joint probability contours for the 
determination of $w_0$ and $w'$ without 
calibration error and with two values of the calibration 
error in the one-temperature model. The increase in uncertainty is quite small
due to the use of multiple filters that correct for calibration variations between them in
the same way as for extinction caused by dust.}
\label{fig:1T}
\end{figure}

For the one-temperature model, Fig.~\ref{fig:1T} shows the 
68\% confidence level contours in the $w_0$-$w'$ plane for the case 
without calibration errors, with a 1\% error in $T$, and
with a 10\% error in $T$.  Even allowing for a 10\% uncertainty 
in $T$ results in a very small
systematic error in the cosmological parameters. This unintuitive 
result can be explained by noting that 
in the Rayleigh-Jeans region a miscalibrated temperature
corresponds mostly to a change in overall flux scale, which is absorbed in ${\cal
M}$, and then the tilt, or color in astronomical terms, is treated by the
extinction correction for flux differences between wavelengths.
For a model with a single 
free parameter, $T$, the correction is almost perfect. 

\begin{figure}
\includegraphics[width=84mm]{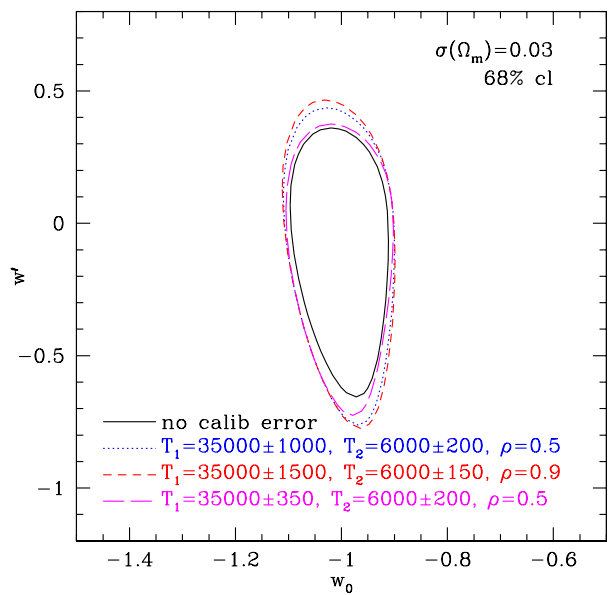}
\caption[]{Joint probability contours for the 
determination of $w_0$ and $w'$ without
calibration error and with several values of the calibration 
errors and correlation in the two-temperature model. Calibration
precision of order 3\% in temperature is sufficient at these temperatures
to keep parameter uncertainties within reasonable
bounds.}
\label{fig:2T}
\end{figure}

Figure~\ref{fig:2T} shows the equivalent contours for the 
two-temperature model.  Now the effect of
the calibration error can be clearly seen. For 3\% uncertainties 
it leads to an increase in the errors of $w_0$ and $w'$ of 
around 20\%, relative to the purely statistical errors. 

The results are fairly insensitive to the correlation coefficient 
$\rho$ until it is nearly one.  Then the calibration reduces to 
the one-temperature model.  A 50\% increase in the magnitude of the 
error in the optical region also does not strongly affect the 
parameter estimation.  So long as the two-temperature model 
realistically approximates the entire calibration procedure, a 
moderate precision at the few percent level
should suffice for calibration to pose only a minor 
contribution to uncertainty in the cosmological parameter 
estimation.  In particular, 3\% errors in both temperatures and 
any degree of correlation between the optical and infrared 
calibration measurements limit the increase
in the overall error to $\la 20\%$ of the result without 
any calibration error. 

Also note that if an optical calibration source 
can be determined to 1\% (as the solar effective temperature 
already is~\cite{ref:susana}), then the error
increase due to the calibration 
systematic uncertainty drops below 10\%.  
And if 
improvements in observations and modeling allow a single body 
like a hot white dwarf to be used to calibrate the whole 
spectrum, then the precision of the calibration can be relaxed 
to 10\%.
\section{Malmquist Bias}\label{sec:malm}
Magnitude-limited searches for supernovae produce data samples 
with a selection effect called Malmquist bias: intrinsically 
brighter supernovae will be preferentially discovered.  As 
supernovae at the fainter end of the
luminosity function fall below the detection threshold, the mean
intrinsic peak magnitude of discovered supernovae is biased lower 
(i.e.~more luminous) than
the mean for the whole population.  The bias increases at higher
redshifts as the apparent magnitudes of the supernova population
grow fainter. This redshift-varying bias will thus enter the
estimation of the cosmological parameters.

Measurements of the intrinsic dispersion in SNe Ia range from 0.25 --
0.35 magnitudes.  However, correcting supernova magnitudes based on 
their light-curve time-evolution (stretch) can reduce the residuals 
to an rms of $\sim 0.10--0.15$ magnitudes.  Note 
that this also leads to a selection effect: supernovae with high 
stretch are more likely to be found -- they remain visible in the 
sky longer and are intrinsically brighter.  However, a well-designed 
survey ensures that the stretch
determination of discovered supernovae is unbiased. Here we consider 
Malmquist bias in the stretch-corrected luminosity function.

In principle, this error can be corrected for if the detection
efficiency and intrinsic luminosity function are known.  But the
luminosity function should vary with the star formation history of the
universe.  Malmquist bias can thus stem from subtle systematic
magnitude shifts arising from uncorrected population evolution of the
progenitor systems.  A precise bias correction requires well
determined luminosity functions over the redshift span of interest.  
One way around this is by brute force: 
the detection threshold can be set much fainter than supernovae at the
highest targeted redshift.

To illustrate the effect of Malmquist bias, we simulate light curves
generated by a space survey with a SNAP-class telescope and a ground
survey with a DMT-class (6.5m, e.g. LSST~\cite{ref:LSST}) telescope
(augmenting the imager with 
an additional near infrared camera) sited at Mauna Kea.  Ground 
observations are taken in the fiducial SNAP filter set with eight-hour 
observations.  The
observing cadence is four observer-frame days.  Mauna Kea weather
condition statistics are from \cite{ref:ESPAS:MK} and the seeing
distribution compiled at Subaru is used~\cite{ref:SubaruSeeing}.  
We include host-galaxy
extinction with an absorption distribution based on \cite{ref:hatano98}. 

\begin{figure}
\includegraphics[width=84mm]{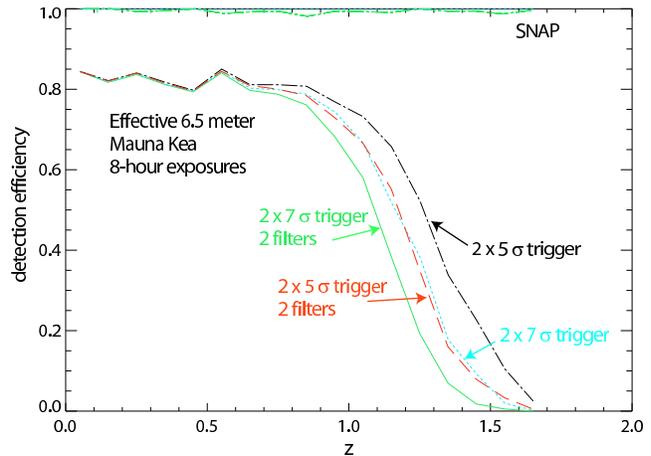}
\caption{The efficiencies for detection of supernovae before maximum light
for a SNAP-class space survey and an effective 6.5 meter telescope at
Mauna Kea taking 8 hour exposures with a 4-day cadence.  Only
supernovae with host-extinction $A_V<0.5$ are included in the
sample. The 0.85 ground efficiency at low redshift is predominantly
due to lost nights due to weather.  The slightly jagged nature of the
curves is due to the finite number of supernova realized in the
simulation.
\label{trigger:fig}}
\end{figure}

We consider four simple detection triggers.  The first two mimic
triggers used in current ground searches and demand at least two
points with signal-to-noise $S/N>5$ (7) in a single passband.  The
second two triggers require significant signal, $S/N>5$ (7), for two
points in two passbands. This is important for the use of color 
time-evolution to distinguish SNe Ia from other transients.  We 
require discovery before
maximum light to allow spectral observations at peak brightness.  The
detection efficiencies for these triggers are shown in
Fig.~\ref{trigger:fig}.  The space mission discovers supernovae with
almost perfect efficiency.  The ground search suffers a
fundamental level of inefficiency which is independent of intrinsic
supernova luminosities since it is due to lost nights from poor
weather.  Efficiency from the ground suffers a further drop-off at
redshifts $z>0.9-1.2$, depending on the trigger.

\begin{figure}
\includegraphics[width=84mm]{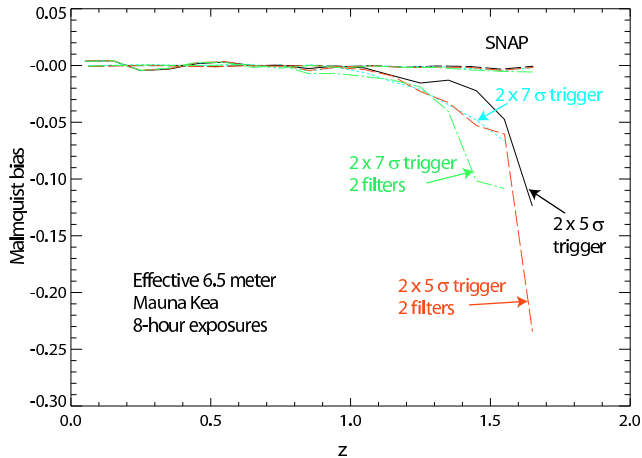}
\caption{Malmquist bias for the four triggers for the space
and ground-based surveys.  Depending on the exact form of the trigger,
the ground detection thresholds drop beginning at $0.9 <z < 1.3$.  The space
mission shows negligible Malmquist bias over the redshift range of
interest.  The small fluctuations are due to the finite
number of supernovae used in calculating the detection
efficiency.
\label{malm:fig}}
\end{figure}

The Malmquist biases induced by these triggers in space and ground
missions are shown in magnitudes in Fig.~\ref{malm:fig}. Depending 
on the exact form of the trigger, the Malmquist bias on the ground 
grows 
beginning 
at $z=0.9-1.3$.  The more rigorous triggers with higher signal-to-noise 
thresholds or requiring detection in two bands have shallower  
redshift reach.  From space, the bias remains $<0.01$ mag for all 
triggers over the full redshift range.

We calculate the effect of this Malmquist bias on the ground searches
described in \S\ref{sec:survey}.  As a representative example, we
consider the results from a trigger consisting of two $S/N>7$ points
in a single filter.  To illustrate the bias that occurs, 
we calculate the
statistical errors and bias in the fit parameters assuming no
irreducible systematic error
for three ground surveys
with $z_{max}=0.7, 1.0$, and 1.3.  For the two shallower surveys,
the level of Malmquist bias is very low (cf.~Fig.~\ref{malm:fig})
and produces a bias that is small compared to the statistical errors.
For the deepest survey, there is a significant 0.04 magnitude
bias at $z_{max}=1.3$ that propagates into a significant component
of the error budget; the bias in the measured cosmological parameters
is now comparable to the pure statistical errors.  There is
little point in obtaining better statistics 
when the systematic errors dominate.
Unless Malmquist bias can otherwise be eliminated or corrected for, deeper
ground searches are thus fundamentally limited.

Note that this analysis only addressed the issue of Malmquist 
bias; there are difficulties posed to ground-based surveys by 
incompletely sampled lightcurves, lack of extinction 
measurements as the supernova flux redshifts into the infrared, 
atmospheric emission, etc.
%
\section{Gravitational Lensing}
\label{sec:lensing}
Another effect on the measured flux of a supernova involves 
magnification or (more frequently) demagnification by weak 
gravitational lensing from intervening mass 
distributions\footnote{Strong gravitational lensing, causing 
multiple images, is rare, affecting roughly one out of every 
thousand objects.  Strong lensing of supernovae can be quite 
interesting and useful for cosmological parameter determination 
\cite{ref:SGL} but it will not significantly affect 
the Hubble diagram.}. Since this conserves flux, the mean effect 
on supernovae at any redshift is null.  However, the finite number 
of supernovae per redshift bin does not sample the entire 
magnification distribution, leading to a slight bias in the 
results from the Hubble diagram analysis. But 
this small bias can be determined from the data themselves~\cite{ref:ariel3},
independently of any model of gravitational lensing,
and hence corrected, leaving only an additional systematic
uncertainty coming from the limited statistical precision of the 
correction itself. 

For weak lensing caused by large-scale structure, the rms of the shift 
is only a few percent and the mean shift can be averaged below 1\% with only a 
couple dozen supernovae per redshift bin \cite{ref:dalal,ref:holzlinder}. 
Here we consider the remaining case of lensing by compact objects, 
which generally gives a broader distribution of magnifications, 
following the method of~\cite{ref:ariel3}.  They show that the
distribution of the biases in magnitude can be adequately 
parameterized by a functional form with three free parameters:
\begin{eqnarray}
f(m) &=& \exp{\left(-\frac{\left(m-m_0\right)^2}{2\sigma^2}\right)}\ , 
\ \ \  m\geq m_c    \nonumber\\
f(m) &=& \exp{\left(-\frac{\left(m-m_0\right)^2}{2\sigma^2}\right)} \nonumber \\
&+& b\cdot |m-m_c|\cdot 10^{s\cdot m}\ ,\ \ \ m\leq m_c \ , 
\end{eqnarray} 
where $m$ is the bias after lensing, and the distribution takes into
account the intrinsic dispersion of supernova magnitudes. The 
distribution represents a 
Gaussian with an extra tail toward demagnification. The constants 
$s$ and $m_c$ are set to 2.5 and 0 respectively.  The three free 
parameters, $m_0$, $\sigma$ and $b$, will in general depend 
on the redshift and the assumed fraction of intervening mass 
in compact objects, $f_p$.  We take $f_p=0.2$, which should 
give an upper bound on any plausible effect. 

For a given redshift $z$ one can determine~\cite{ref:ariel2} 
the set $(m_0,\sigma,b)$ that parameterizes the $m$ 
distribution~\footnote{Note that the distribution depends on $z$, with
$\sigma$ generally increasing with $z$: hence its possible influence on
cosmological parameter determination.}.
We then generate a Monte Carlo 
sample of $N(z)$ supernovae with the 
distorted $m$ distribution. This distribution
can be measured in real data by shifting the magnitudes of all 
supernovae in the bin to the value they would have had if their
redshift had been that of the center of the bin. For this, one has 
to assume a certain cosmology, but the result is fairly
insensitive to this for reasonable bin widths, and by iterating 
the procedure after the cosmology fit all such dependency can be 
suppressed.

Once the $m$ distribution (in real data or through Monte Carlo) 
is obtained, a fit is made in order to measure the
parameters of the distribution. Call the result $(m_0',\sigma',b')$. 
Each parameter will have been measured
with a certain error. The central value of the distribution 
corresponds to the net bias in magnitude that has to be
corrected. Monte Carlo is used in order 
to compute the uncertainty in this central value as a function 
of the uncertainties in $(m_0',\sigma',b')$. This procedure has 
been implemented for $z=0.5,1.0,1.5$ and $N(z)$ from the fiducial 
distribution in Table~\ref{tab:z}.
The uncertainties after the corrections have been
found to be 0.017, 0.018 and 0.021 mag for the three redshifts, 
therefore in line with the assumptions made in previous 
sections of a systematic error around 2\%.  A more 
realistic compact object fraction, $f_p < 0.1$, would give a
significantly smaller effect. 
\section{Evolution}
\label{sec:evolution}
In the previous sections we have addressed systematic 
uncertainties arising from the detectors, the observation 
strategy, and the light propagation.  A potentially serious 
systematic error lies in the supernovae themselves.  While 
supernovae have no direct influence from the cosmic time, 
i.e.~they don't ``know'' what redshift they are at 
\cite{ref:coping}, more subtle ``evolutionary'' effects can enter 
in the form of population drift. 

For example, suppose the metallicity of a supernova had an influence 
on its absolute luminosity and supernovae at higher redshifts  
systematically had a lower metallicity than nearby supernovae. 
In this case, if the survey did not recognize and correct for 
this trend then the resulting Hubble diagram would be biased 
and the parameters resulting from the cosmology fit inaccurate. 

Two approaches to this problem can be considered.  Both require 
the survey to acquire and use the wealth of observational 
information that each supernova provides in the form of its 
time series of flux and its energy spectrum.  Then the key 
assumption is that the state of the supernova is adequately 
described by these detailed measurements, that supernovae 
with the same lightcurve and spectrum indeed have the same 
absolute magnitude, that there are no ``hidden variables''.  

One approach is to carry out comprehensive modeling of 
Type Ia supernova explosions and radiation transport, 
examining numerically the effects of varying progenitor 
characteristics.  The underlying physics proves remarkably 
robust and the final state possesses an extraordinary degree 
of ``stellar amnesia'' to the initial conditions \cite{ref:hoeflich03}. 
The remaining small variations in the light curves and spectra 
may even provide a tightening of the standard candle nature 
through 
a secondary correction parameter 
beyond stretch. 

However we adopt only the purely empirical path of examining 
detailed observations of a wide range of supernovae.  Any modeling 
is used solely to suggest features that might be of interest in 
the lightcurve and spectrum, and to serve as a subsidiary 
crosscheck.  The Nearby Supernova Factory project~\cite{ref:snf} will provide 
invaluable information on identifying secondary characteristics 
influencing the magnitude.  Such a survey samples the breadth of 
galactic environments and physical conditions present at any 
redshift, e.g.~the range of metallicities.  However as mentioned 
above, there could be a population drift --a change in the 
proportion of environments at the different redshifts.  

While analyses of supernovae with an average redshift of 
$z\approx0.5$ \cite{ref:sullivanellis} show no sign of a systematic 
trend (within the current precision)
with galaxy type or location of the supernova within the 
galaxy, we need to guard against the possibility at higher 
redshift.  This leads to the program of ``like-to-like'' 
comparisons~\cite{ref:systSCP,ref:coping}.  
One categorizes the supernovae based on the 
detailed observations into subsamples with similar intrinsic 
characteristics, e.g.~ratios of peak-to-late-time light-curve 
magnitude, ultraviolet properties, line ratios, etc.  Then 
these narrow subclasses are compared across different redshifts, 
taking that like spectral and flux characteristics 
imply like intrinsic magnitudes --with sufficiently comprehensive 
measurements there is nowhere for a change to hide.  This 
procedure, we emphasize, is purely empirical and does not depend 
on any theoretical model of the luminosity or of evolution. 
The mantra is that like supernovae at different redshifts give 
a clear view of cosmology, and different supernovae at like 
redshifts alert us to intrinsic systematics. 

To test this approach, we divided the total sample of SNe in 
up to ten subsamples, each 
with roughly the same number of SNe and similar distribution in $z$, 
allowing a different intrinsic magnitude (i.e.~$\calm$) for each 
subsample in the fit. The fit returned the same cosmological 
parameters as the fit with just one $\calm$, with a negligible increase in 
their statistical error. 

We then allow the subsamples to have different redshift distributions, 
mocking up population drift.  The 2000 supernovae are divided into 
three subsamples of roughly the same size, one of them with the 
number of supernovae rising linearly with $z$, another one decreasing
linearly and the third one flat, so that the total sample is uniform 
in redshift.  When a different value for the intrinsic magnitude is 
allowed for each subsample, one finds a small (less than 4\%) 
increase in the uncertainty in $w_0$, relative to the 
case of a single sample of 2000 SNe uniformly distributed in $z$ 
and with a single intrinsic magnitude for all the SNe. 
An even smaller increase (below 2\%) is seen for $w'$, while 
essentially no change is observed for $\Omega_m$.  Thus 
comprehensive data collection, including spectra, together with 
like vs.~like analysis, appear to provide a robust solution for 
potential evolutionary systematics. 
\section{Conclusion}
Precision cosmological observations offer the hope to uncover 
essential properties of our universe, including the nature of 
the dark energy that causes the present accelerating expansion 
and could determine its fate.  But hand in hand with these 
advances must go understanding of the systematic effects that 
could mislead us.  We have presented in some generality as 
well as some detail several possible sources of systematic 
uncertainty for the Type Ia supernova method of mapping the 
cosmological distance-redshift relation. 

In every case we have shown that the residual uncertainties 
after using detailed measurements of the light curves and 
spectra, when bounded below 0.02 mag, do not significantly 
interfere with the goal of accurate estimation of the matter 
and dark energy densities, $\sigma(\Omega_\Lambda) = 0.03$, 
the dark energy equation of state 
today, $\sigma(w_0)=0.07$, and a measure of its time variation, 
$\sigma(w')=0.3$. 

This supports with analytic, numerical, and Monte Carlo 
simulations the conclusion that a well designed satellite survey 
of about 2000 Type Ia supernova, observed out to redshift 
$z=1.7$, with complete lightcurve characterization and a 
spectrum for every supernova, can succeed in answering some 
of the most fundamental questions about our universe and 
physics. 

No other cosmological probes, promising though they might 
appear, have yet addressed the crucial question of systematics 
with the same degree of rigor.  When this is established they 
may well offer valuable complementarity.  For the supernova 
method, we note that the combined optical and near infrared 
observations and redshift reach to $z=1.7$ are critical 
elements in reducing the impact of systematic uncertainties.  
\section*{Acknowledgments}
We are grateful to numerous people for discussing the many varied 
aspects of astrophysics and instruments that enter into this work. 
We would especially like to acknowledge Greg Aldering, David Branch, 
Susana Deustua, Peter H{\" o}flich, Dragan Huterer, 
Michael Lampton, Michael Levi, Stuart Mufson and Saul Perlmutter.  
We would like to thank Ariel Goobar and Edvard M\"{o}rtsell for making
their results~\cite{ref:ariel3} available to us before publication.
We thank Gary Bernstein for his careful reading of the manuscript and his
insightful comments. 
This work was supported 
by the Director, Office of Science, US Department of Energy, under contracts
DE-AC03-76SF00098 (LBNL) and DE-FG02-91ER40661 (Indiana).
RM is partially supported by the US National Science Foundation under agreement
PHY-0070972. EVL thanks the KITP Santa Barbara for hospitality during  
part of the paper preparation. 
\appendix
\section{Fisher Matrix Analysis of Systematic Magnitude Offsets}
A constant level 
of magnitude offset can be absorbed wholly into the $\calm$ 
parameter, leaving the cosmological parameters unaffected. We use the 
following proof as an illustration of the 
Fisher matrix method. 

The Fisher, or information, matrix method of error estimation 
approximates the parameter likelihood surface by a paraboloid in 
the vicinity of its maximum.
As long as the parameter 
errors are small, the Fisher method gives an excellent approximation 
to a full maximum likelihood analysis.  The 
Fisher matrix relates the observables, in this case the set of 
supernova magnitudes $m(z)$, to the parameters $\theta=\{\om,w_0,w', 
\calm\}$ through the sensitivities $\partial m/\partial\theta$: 
\beq 
F_{ij}={1\over\sigma_m^2}\int dz\,N(z)\dmthi \dmthj\, ,\label{fish}
\eeq 
where $\sigma_m = 0.15$ and $N(z)$ is the number of SNe in a redshift bin around $z$.
The error, or covariance, matrix is the inverse of this, so for 
example $\sigma^2(w_0)=(F^{-1})_{w_0w_0}$. 

External or prior information is incorporated simply by adding the 
Fisher matrices.  The simplest example is a Gaussian prior on a 
single parameter, say $\om$, which corresponds to adding an information 
matrix empty save for a single entry in the appropriate diagonal space;  
if prior information determines $\om$ to $\pm 0.03$ then 
the matrix entry is $1/(0.03)^2$.  Rules of matrix algebra 
allow one to calculate how such a prior affects all the entries in 
the covariance matrix, i.e.~the parameter estimation errors. 

A systematic offset in the observed magnitude similarly propagates 
into the results, in the form of a bias giving a best fit parameter 
value different from the input cosmological model (because this shifts 
locations on the likelihood surface it will also have a small effect on 
the statistical
part of the errors).  Denoting the offset 
as $\dl m(z)$, matrix algebra provides the bias relation 
\beq 
\dl\theta_i=F^{-1}{}_{ij}{1\over\sigma_m^2}\int dz\,N(z)\,\dl m(z)\dmthj\,, 
\label{bias}\eeq 
where summation over repeated indices is implied.  This allows 
straightforward calculation of the induced bias, within the 
Fisher formalism.  Note that this looks similar to the Fisher matrix 
expression (\ref{fish}) but only contains a single sensitivity factor. 

However, the nuisance parameter $\calm$ is purely additive and so has 
sensitivity $\partial m/ \partial\calm=1$.  Thus $F_{\calm\theta}$ has 
only one apparent derivative factor, like the bias expression.  Indeed 
if we separate 
out the redshift independent part of the systematic, $\dl m(z)=\dl m_0 
+dm(z)$, then the term containing the constant offset reduces to 
\beq 
\dl\theta_i=F^{-1}{}_{ij}F_{j\calm}\,\dl m_0=\dl m_0\,\dl_{i\calm}.
\eeq 
So the constant part of the offset systematic only causes a bias in the 
nuisance parameter $\calm$ and does not affect the cosmological 
parameters.  One could choose this to represent the mean offset, or 
to remove a constant from the offset such that the magnitude systematic 
is defined to be zero at zero redshift. 

\begin{thebibliography}{99}
%
\bibitem[Akerlof et al.~2003]{ref:omnibus} 
Akerlof, C.~ et al., 2003, in preparation

\bibitem[Aldering 2002]{ref:SubaruSeeing} 
Aldering, G., 2002, SNAP internal memo

\bibitem[Aldering et al.~2002]{ref:snf} 
Aldering, G.~et al., 2002, \procspie, 4836; \\
http://snfactory.lbl.gov/spie\_2002.pdf

\bibitem[Allen, Schmidt \& Fabian, 2002]{ref:ASF}
Allen, S.W., Schmidt, R.W., \& Fabian, A.C., 2002, MNRAS, 334, L11
[arXiv:astro-ph/0205007]

\bibitem[Amanullah, M\"{o}rtsell \& Goobar 2003]{ref:ariel3}
Amanullah, R., M\"{o}rtsell, E., \& Goobar, A., 2003, \aap, 397, 819

\bibitem[Bohlin 2002]{ref:bohlin02}
Bohlin, R., 2002, Proc. of 2002 HST Calibration Workshop, 97 

\bibitem[Branch et al.~2001]{ref:coping}
Branch, D., Perlmutter, S., Baron, E., \& Nugent, P., 2001, arXiv:astro-ph/0109070 

\bibitem[Brax \& Martin 1999]{ref:braxm}
Brax, P.~\& Martin, J., 1999, PLB 468, 40

\bibitem[Cardelli, Clayton \& Mathis 1989]{ref:cardelli}
Cardelli, J.~A., Clayton, G.~C., \& Mathis, J.~S., 1989, \apj, 345, 245

\bibitem[Carroll 2001]{ref:carroll}
Carroll, S.~M., 2001, arXiv:astro-ph/0107571,\\
 http://pancake.uchicago.edu/\~{ }carroll/preposterous.html

\bibitem[Commins 2002]{ref:commins}
Commins, E.~D., 2002, SNAP internal note

\bibitem[Dalal et al.~2003]{ref:dalal} 
Dalal, N., Holz, D.~E., Chen, X., \& Frieman, J.~A., 2003, \apjl, 585, L11 

\bibitem[Deustua 2002]{ref:susana}
Deustua, S.~E., 2002, private communication

\bibitem[Essence 2003]{ref:ess}
Essence, 2003, http://www.ctio.noao.edu/essence 

\bibitem[Frieman et al.~2003]{ref:fhlt} 
Frieman, J.~A., Huterer, D., Linder, E.~V., \& Turner, M.~S., 2003, \prd, 67, 083505 
[arXiv:astro-ph/0208100] 


\bibitem[Hatano, Branch \& Deaton 1998]{ref:hatano98}
Hatano, K., Branch, D., \& Deaton, J., 1998,
\apj, 502, 177
[arXiv:astro-ph/9711311]

\bibitem[H\"{o}flich et al.~2003]{ref:hoeflich03}
H\"{o}flich, P., Gerardy, C., Linder, E.~\& Marion, H., 2003, 
arXiv:astro-ph/0301334, in ``Stellar Candles'', 
eds.~Gieren et at., Lecture Notes in Physics

\bibitem[Holz \& Linder 2003]{ref:holzlinder}
Holz, D.~E.~\& Linder, E.~V., 2003, in preparation

\bibitem[Huterer \& Turner 2001]{ref:huttur}
Huterer, D.~\& Turner, M.~S., 2001, \prd, 64, 123527 

\bibitem[Kim, Goobar \& Perlmutter 1996]{ref:KGP}
Kim, A., Goobar, A., \& Perlmutter, S., 1996, PASP, 108, 190



\bibitem[Lampton 2002]{ref:lampton}
Lampton, M.~L., 2002, SNAP internal note

\bibitem[Linder \& Huterer 2003]{ref:linhut} 
Linder, E.~V.~\& Huterer, D., 2003, \prd, 67, 081303 [arxiv:astro-ph/0208138]

\bibitem[LSST 2003]{ref:LSST}
LSST, 2003, http://www.dmtelescope.org

\bibitem[Minuit 2002]{ref:minuit}
Minuit, 2002, http://wwwinfo.cern.ch/asdoc/minuit/minmain.html

\bibitem[M\"{o}rtsell 2002]{ref:ariel2}
M\"{o}rtsell, E., 2002, private communication. See also~\cite{ref:ariel3}

\bibitem[NAG 2002]{ref:NAG}
NAG, 2002, http://anaphe.web.cern.ch/anaphe/gemini.html

\bibitem[Oguri, Suto \& Turner 2003]{ref:SGL}
Oguri, M., Suto, Y., \& Turner, E.~L., 2003,
\apj, 583, 584 [arXiv:astro-ph/0210107]

\bibitem[Perlmutter \& Schmidt 2003]{ref:systSCP} 
Perlmutter, S.~\& Schmidt, B.~P., 2003, arXiv:astro-ph/0303428, to appear in
``Supernovae and Gamma Ray Bursts'', ed.~K.~Weiler, Lecture Notes in
Physics

\bibitem[Perlmuter et al.~1999]{ref:SCP} 
Perlmutter, S.~et al., 1999, \apj, 517, 565 
 
\bibitem[Phillips et al.~1999]{ref:phillips99}
Phillips, M.~M.~et al., 1999,
\aj, 118, 1776
[arXiv:astro-ph/9907052]

\bibitem[Riess, Press \& Kirshner 1996]{ref:rpk96}
Riess, A.~G., Press, W.~H., \& Kirshner, R.~P., 1996,
\apj, 473, 88
[arXiv:astro-ph/9604143]

\bibitem[Riess et al.~1998]{ref:High-Z} 
Riess, A.~G.~et al., 1998, \aj, 116, 1009 

\bibitem[Sarazin 2002]{ref:ESPAS:MK}
Sarazin, M., 2002, {\em ESPAS Site Summary Series: Mauna Kea}, ESO report

\bibitem[SNAP 2003]{ref:snap}
SNAP, 2003, http://snap.lbl.gov

\bibitem[Spergel et al.~2003]{ref:CMB} 
Spergel, D.~N.~et al., 2003, arXiv:astro-ph/0302209 

\bibitem[Sullivan et al.~2002]{ref:sullivanellis}
Sullivan, M.~et al., 2002, arXiv:astro-ph/0211444 

\bibitem[Tripp \& Branch 1999]{ref:Tripp:1999}
Tripp, R.~\& Branch, D., 1999,
Correction for Type Ia Supernovae,''
\apj, 525, 209
[arXiv:astro-ph/9904347]

\bibitem[Weller \& Albrecht 2001]{ref:welal} 
Weller, J.~\& Albrecht, A.~, 2001, \prl, 86, 1939 

%
\end{thebibliography}
\end{document}